\documentclass[prl,twocolumn]{revtex4}

\usepackage{amsmath}
\usepackage{amssymb}
\usepackage{amsfonts}
\usepackage{color}
\usepackage{graphicx}

\newcommand{\m}{\mathrm}

\begin{document}

\title{Graphene optomechanics realized at microwave frequencies}

\author{X.~Song$^1$}
\author{M.~Oksanen$^1$}
\author{J.~Li$^1$}
\author{P.~J.~Hakonen$^1$}
\author{M.~A.~Sillanp\"a\"a$^{1,2}$}
\thanks{Mika.Sillanpaa@aalto.fi}
\affiliation{$^1$O.~V.~Lounasmaa Laboratory, Low Temperature Laboratory, Aalto University, P.O. Box 15100, FI-00076 Aalto, Finland. \\
$^2$Department of Applied Physics, Aalto University School of Science, P.O. Box 11100, FI-00076 Aalto, Finland}

\begin{abstract}
Cavity optomechanics has served as a platform for studying the interaction between light and micromechanical motion via radiation pressure. Here we observe such phenomena with a graphene mechanical resonator coupled to an electromagnetic mode. We measure thermal motion and back-action cooling in a bilayer graphene resonator coupled to a microwave on-chip cavity. We detect the lowest flexural mode at 24 MHz down to 50 mK, corresponding to roughly mechanical 40 quanta, representing nearly three orders of magnitude lower phonon occupation than recorded to date with graphene resonators.
\end{abstract}

\maketitle

Graphene can be considered as an ultimate material for studying the quantum behavior of the motion of micromechanical resonators. It is lightweight, and therefore the zero-point motion $x_{\m{zp}} = \sqrt{\hbar/2 m \omega_m}$ of a particular mode of frequency $\omega_m$ and mass $m$ is unusually large. It is stiff, such that the frequency is high for a given mass, allowing for approaching the quantum limit in dilution refrigerator temperatures. Moreover, graphene resonators can portray appreciable nonlinearity at single-quantum level. These properties would make graphene attractive for the use in cavity optomechanics experiments.

In the field of cavity optomechanics \cite{OptoReview2014}, remarkable findings have been made during the last ten years. These include, for example, ground state cooling of a mechanical mode \cite{Teufel2011b,AspelmeyerCool11}. Briefly, an optical or more generally electromagnetic cavity mode with a movable mirror or boundary condition allows to couple the confined photons and the motion by means of the radiation pressure exerted by the photons.

Photothermal interaction between the motion of a mechanical resonator made with graphene \cite{ParpiaGraph,Bachtold2008GR,Hone2010,Kim2013GR} and light was observed quite recently \cite{ParpiaOpto}. These authors had a on-chip trench covered by graphene, thus forming the cavity. However, since the graphene end mirror is 98 \% transparent, the $Q$-value was much less than one, and canonical radiation pressure phenomena could not be observed.

In contrast to optical frequencies, graphene naturally lends itself to microwave-frequency cavities as it is opaque at these frequencies. The driven motion of a graphene mechanical resonator was observed by coupling to an off-chip tank circuit \cite{GrapheneStamp}, however, the coupling was not sufficient to observe graphene thermal motion not to mention cavity back-action.

In this work we make the first demonstration of radiation pressure effects on the motion of graphene in the context of cavity optomechanics. Cooled in a dilution refrigerator, we study the lowest drum-like flexural mode of a bilayer graphene resonator at $\omega_m/2\pi = 24$ MHz, coupled to a high-$Q$ microwave cavity resonator. We observe thermal motion down to 70 mK, furthermore, we carry out cavity back-action sideband cooling to further dampen the effective mode temperature down to 50 mK. A thermal transport model is developed to explain the findings.

\begin{figure}[ht]
 \includegraphics[width=0.95\linewidth]{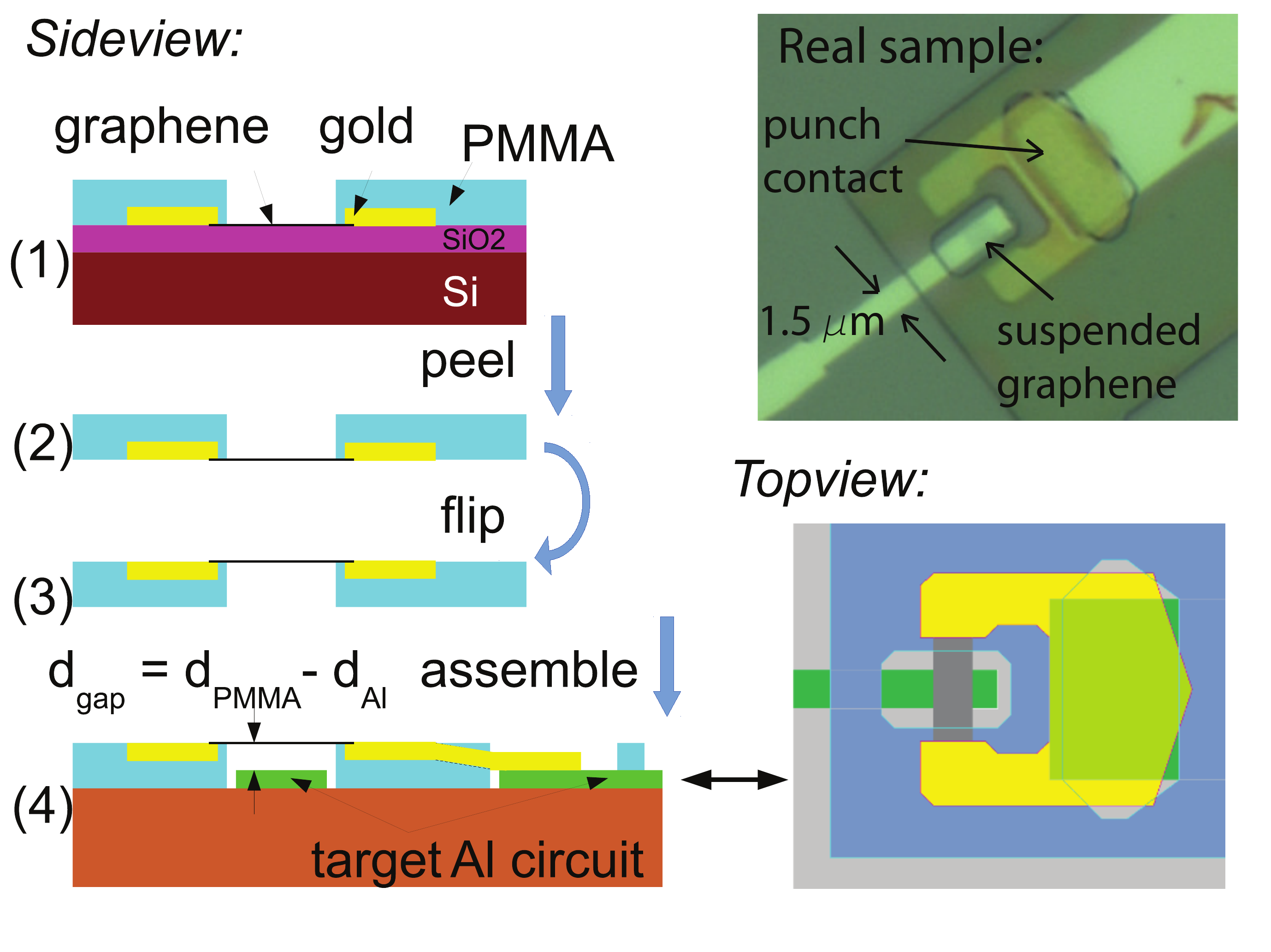}
  \caption{\emph{Device fabrication}. Gold contacts 30 nm thick are deposited on the previously located graphene pieces by e-beam lithography (EBL) and lift-off.  Then PMMA is spun on the chip again and a 2nd EBL step is applied (1) to define patterns such as stamps and holes in it. The whole PMMA membrane is then peeled off from the initial substrate (2) by etching away the SiO$_2$ in 1\% HF solution. We pick up one of such stamps, flip and transfer it (3-4) over the aluminum electrode to form a parallel plate capacitor.}
  \label{fig:fab}
\end{figure}

\begin{figure*}[htp]
 \includegraphics[width=0.85\linewidth]{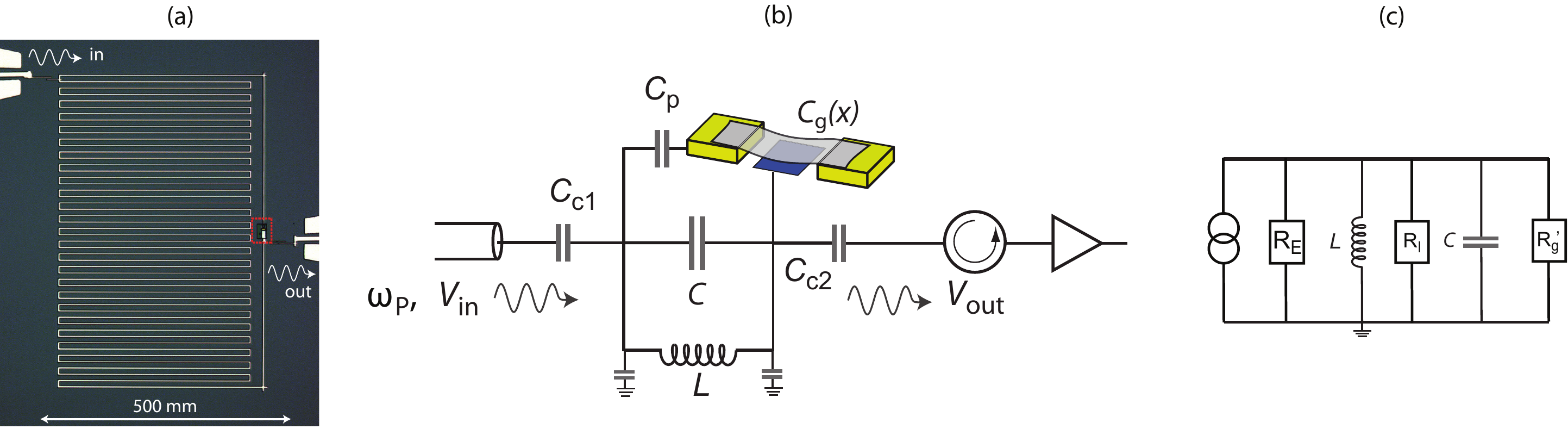}
  \caption{\emph{Schematic of the experiment}. (a) The microwave cavity resonant at 7.8 GHz. White is Al film 80 nm thick, dark is Si substrate. (b) Scheme of the transmission measurement through the cavity. The output signal from the cavity is amplified at 4 K stage by a low-noise microwave amplifier, and detected at room temperature by a spectrum analyzer. (c) Equivalent circuit of the cavity including graphene.}
  \label{fig:schema}
\end{figure*}

The interaction of a mechanical resonator with a microwave-regime electrical cavity is similar to the radiation pressure force; displacement $x$ affects the cavity frequency $\omega_c$. This is naturally described in terms of an equivalent capacitance $C$ of the cavity, and a movable capacitance $C_g$ of the conductive mechanical part. The coupling energy, that is, how much the zero-point motion $x_{\m{zp}}$ displaces the cavity frequency, is given by $g=(\omega_c/2C)(\partial C_g/\partial x) x_{\m{zp}}$.

Since for a plate capacitor $\partial C_g/\partial x \propto x^{-2}$, a narrow vacuum gap $d$ is instrumental for obtaining a high $g$. Although $g$ also grows with surface area of the membrane, the dependence is linear and thus our design emphasizes a small gap. A small size in the low micron-range gives the benefit that the frequency is higher (tens of MHz) and quantum limit more easily attainable than with a large-size design.

We have developed a technique to fabricate narrow gaps $d \sim 70$ nm for a membrane about 2.5 $\mu$m long and 1.5 $\mu$m wide in a bridge geometry. First, high-quality tape-exfoliated bilayer graphene pieces are located on a silicon substrate covered with 275 nm thick SiO$_2$ and confirmed with Raman spectroscopy.  After the lithography and transfer process illustrated in Fig.~\ref{fig:fab}, the other end of the gold contact is punched down well in contact to the other aluminum electrode over an area of $\sim 8 \times 3$ $\mu$m. Since we have a good control of the thickness of the aluminum electrode $d_{\m{Al}} = 80$ nm as well as that of the PMMA stamp $d_{\m{PMMA}} = 150$ nm, the vacuum gap is determined as $d = d_{\m{PMMA}} - d_{\m{Al}} = 70$ nm. 

Although the punch from gold to aluminum breaks the Al oxide, a galvanic contact between Au and Al is not established since the Al immediately oxidizes. Hence, the graphene is electrically floating. However, the capacitance between Au and Al satisfies $C_p \gg C_g$ such that the coupling is fully determined by the movable capacitance $C_g(x)$. 
We expect the charge density of the graphene to be low and its resistance thus to be high $R_g \sim 1 ... 10$ k$\Omega$. 


The cavity is a mm-size meandering $\lambda/2$ design made out of aluminum on high-purity silicon chip (Fig.~\ref{fig:schema}a). Since both ends of the cavity are floating, the effective capacitance $C$ in the circuit schematics (Fig.~\ref{fig:schema}b) is very low given the dimensions, $C \simeq 45$ fF.
We can convert the real circuit in Fig.~\ref{fig:schema}b into an equivalent parallel resonator (Fig.~\ref{fig:schema}c) where the graphene resistance is included in $R_g' \simeq 1/(\omega_c C_g)^2 R_g \gg R_g$. With the current parameters $C_g, C_c, C_s \ll C$, the effective capacitance essentially equals $C$. Also, the external and purely cavity losses $R_E, R_I \gg R_g'$.

From the parameters of the graphene and the cavity, we obtain a prediction for the coupling of the lowest flexural mode $g/2\pi \simeq 70$ Hz. However, the data is best fitted (see below) by clearly lower $g/2\pi \simeq 23$ Hz. We attribute this to a tendency of the membrane mode shape to be distorted by non-uniform strain at the clamps.

The measurements are carried out in a dilution refrigerator down to 22 mK base temperature. We measure transmission of near-resonant microwaves through the cavity. In the data discussed here, we apply pump microwave irradiation at the red motional sideband frequency $\omega_{\m{P}} = \omega_c - \omega_m$.  The cavity linewidth below 350 mK is found as $\kappa/2\pi \sim 3$ MHz. This is set by the effective resistance $R_g'$ of the graphene (Fig.~\ref{fig:schema}c), thus reducing the $Q$-value of the cavity. This finding is consistent with about $1-5$ k$\Omega$ total resistance of the graphene.

\begin{figure*}[htp]
 \includegraphics[width=0.75\linewidth]{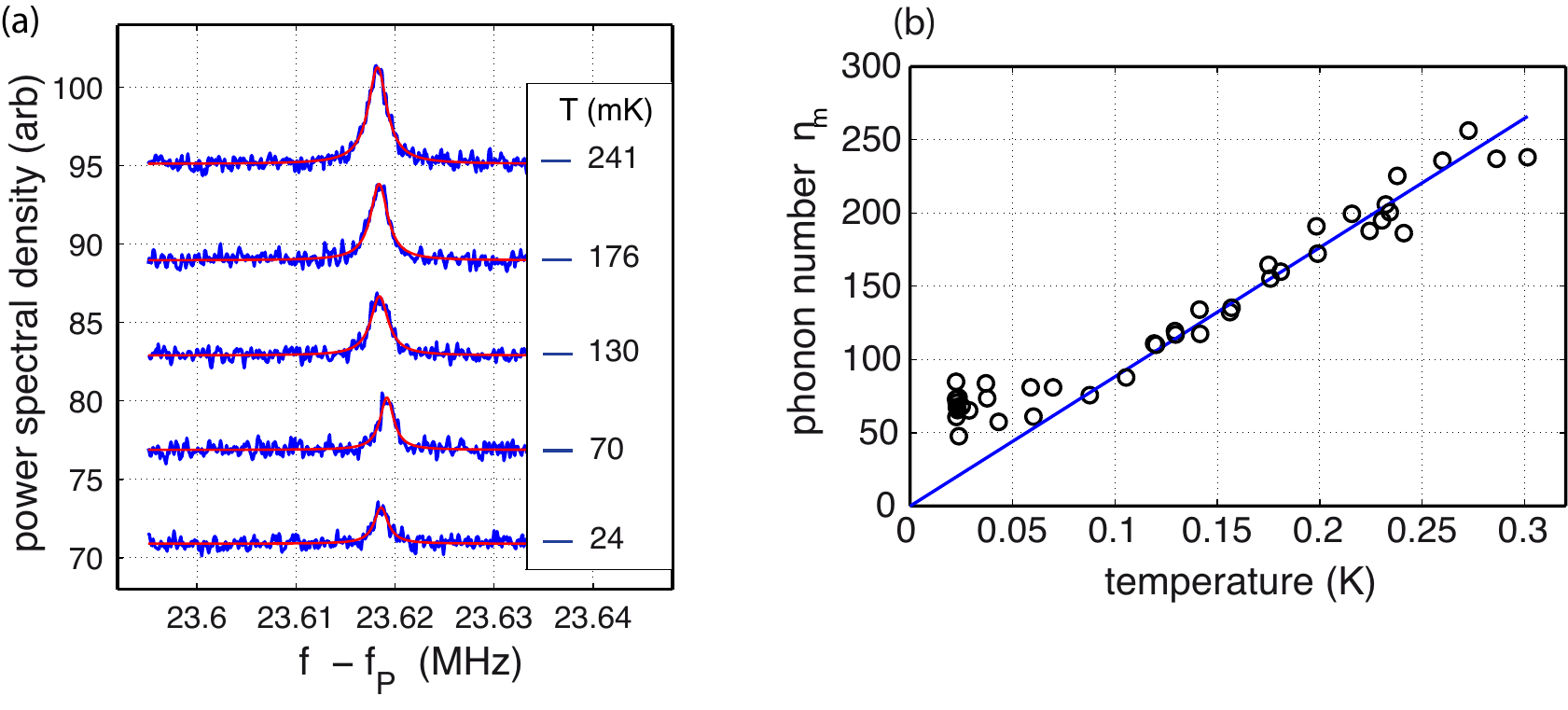}
  \caption{\emph{Thermal motion of graphene}. (a) Spectral density near the motional sideband corresponding to the lowest flexural mode. The curves are shifted vertically for clarity by 6 units. The red lines are Lorentzian fits; (b) Measured phonon number (circles) at different cryostat temperatures. There are several data points nearly on top of each other at 22 mK. The solid line is a least-squares fit to the data above 100 mK. In both (a) and (b), the pump power was about 5 nW, inducing a cavity photon number $n_c \sim 6 \times 10^4$ at $\omega_p$.}
  \label{fig:Tdep}
\end{figure*}

The information on the mechanical resonator appears in the motional sidebands spaced $\pm \omega_m$ about the pump frequency.
In Fig.~\ref{fig:Tdep}a we show results for such a measurement which depict the thermal motion of the lowest mode of the suspended graphene. We used such low pump power that the cavity back-action damping (see below) had a negligible effet. The mechanical $Q$-value at the base temperature was about $15 \times 10^3$.

As clear in Fig.~\ref{fig:Tdep}a, the height of the peaks grows with temperature.
Ideally, the peak area is linearly proportional to cryostat temperature $T_0$, since by equipartition, $1/2 m \omega_m^2 \langle x^2 \rangle= 1/2 k_B T_0$. Since it is very difficult to estimate with reasonable error bars everything affecting the peak height, such as cable attenuation, amplifier gain, or cavity internal losses, we follow the usual practice and calibrate the transduction between peak area and phonon number by relying on the linear temperature dependence, here obeyed above about $T_0 = 70$ mK as seen in Fig.~\ref{fig:Tdep}b. We thus conclude that the lowest flexural mode thermalizes down to 70 mK in this experiment. This represents three orders of magnitude lower phonon occupancy than previous observations of graphene thermal motion, made at room temperature \cite{ParpiaGraph}.

Sideband cooling owing to the radiation pressure interaction has been an extremely powerful and popular tool for studying micromechanical motion \cite{Gigan:2006dn,Kippenberg2009,Schwab2010,Teufel2011b,AspelmeyerCool11,multimode2012}. Here, the pump microwave has to be applied at the red sideband, $\omega_{\m{P}} = \omega_c - \omega_m$. This pump condition is similar to that used in the mere detection of the thermal motion as in Fig.~\ref{fig:Tdep}, but now the pump power is orders of magnitude stronger. 
The sideband cooling process results in an enhanced effective damping experienced by the mechanical mode: $\gamma_{\m{eff}} =\gamma_m + \gamma_{\m{opt}}$, where $\gamma_{\m{opt}}= 4 g^2 n_c/\kappa$. The total damping then gives a reduced phonon occupation: $n_m = \gamma_m n_m^T /\gamma_{\m{eff}}$. Here, the thermal phonon number $n_m^T$ is set by the bath at the temperature $T_{\m{env}}$ according to  $\hbar \omega_m n_m^T = k_B T_{\m{env}}$. Although ideally the bath equilibrates at the cryostat temperature, in the following we find that the microwave heating causes that $T_{\m{env}} \gg T_0$.

\begin{figure*}[htp]
 \includegraphics[width=0.65\linewidth]{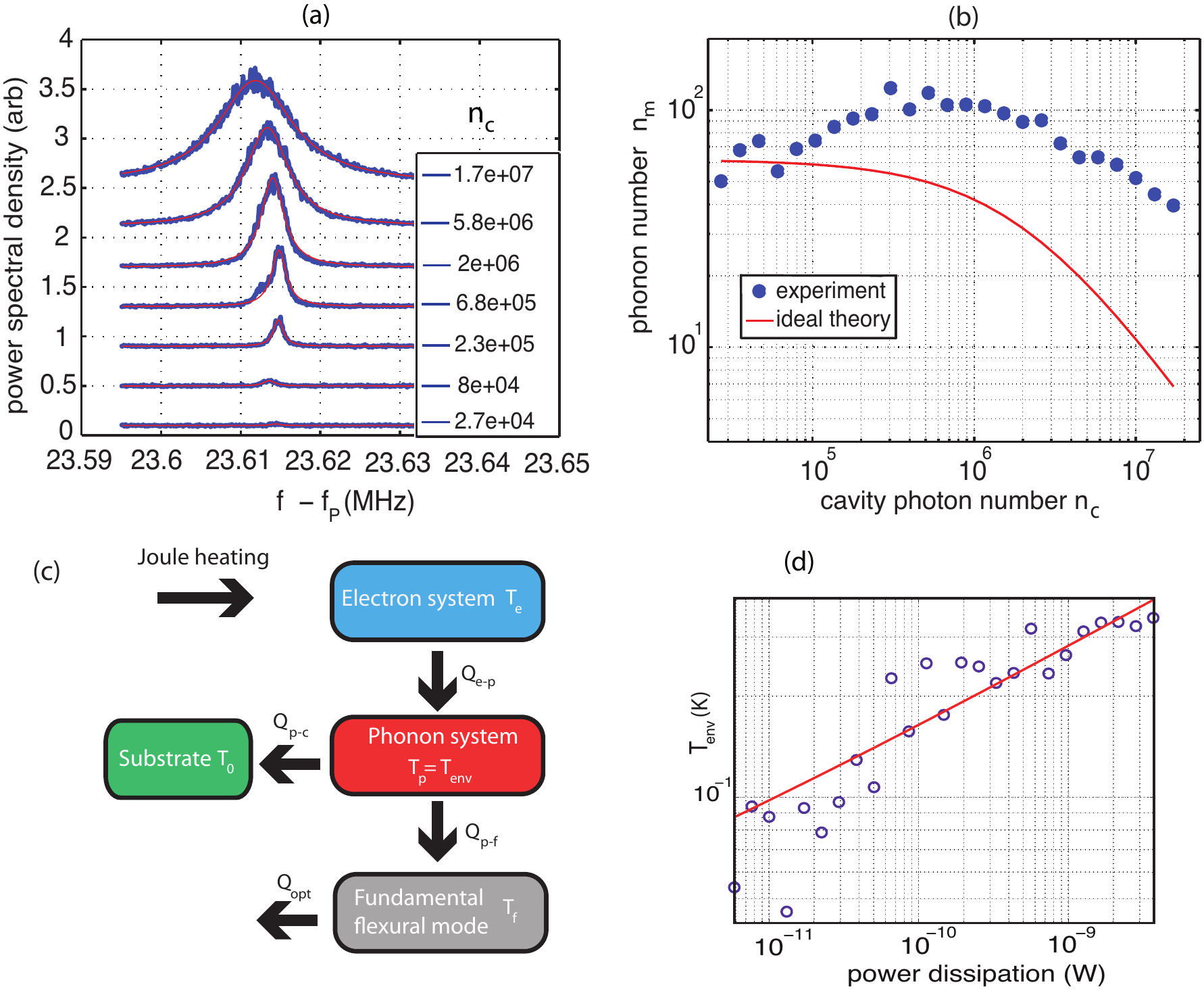}
  \caption{\emph{Cooling and heating}. (a) Thermal motion spectra representing progressing sideband cooling at increasing cavity pump photon occupancies. The curves have been shifted vertically for clarity. The red lines are Lorentzian fits. (b), Phonon number of the graphene mechanical mode versus pump occupancy. The solid line is the ideal behavior. (c), Thermal model used to describe microwave heating. (d), Effective temperature of the environment of the mechanical mode versus power dissipation in the graphene, related to pump occupancy. The solid line is a $T^{4}$ fit.}
  \label{fig:cool}
\end{figure*}

In the measurement, we indeed observe the thermal motion peak becoming broader as the pump power is increased (Fig.~\ref{fig:cool}a), consistent with pure microwave radiation pressure interaction between the cavity and graphene. However, the height of the peaks grow progressively larger. We attribute this to an enhanced temperature of the environment experienced by the graphene mechanical mode. We can first extract the phonon number $n_m$ from the Lorentzians in Fig.~\ref{fig:cool}a, showing progress of the sideband cooling in Fig.~\ref{fig:cool}b. The cooling process clearly deviates from the ideal picture, and shows unusual non-monotonic behavior. Knowing $n_m$ and $\gamma_{\m{opt}}$, we can also deduce the environmental phonon number $n_m^T$  (Fig.~\ref{fig:cool}d). A relatively weak dependence of the temperature on microwave power hints that there is strongly temperature-dependent bottleneck setting the contact of the flexural mode to its bath.



We can describe the mode temperature with a thermal model (Fig.~\ref{fig:cool}c) of the chip at a cryostat temperature $T_0$. We start by supposing the graphene electrons heat by Joule heating by microwave currents. Because the graphene is electrically floating, heat flow out from the electrons $\dot{Q}_{e-p}$ is entirely due to electron-phonon coupling. The phonons are scattered into the substrate across the relatively large contact area, the corresponding heat flow being $\dot{Q}_{p-c}$. We also suppose the thermal environment of the lowest flexural mode is set by the phonon bath which couples via mechanical nonlinearities.
The mode temperature is also affected by the sideband cooling with heat flow $\dot{Q}_{\m{opt}}$. In a typical case here, $\dot{Q}_{\m{opt}} \ll \dot{Q}_{p-c}$.


At low-temperatures $T \ll T_{BG}$, assuming parabolic bands \cite{McCann2006}, the electron phonon coupling in bilayer graphene can be written as \cite{Viljas2010}: $\dot{Q}_{e-p}=A \Sigma (T_e^4-T_{p}^4)$, where $A$ is the graphene area, $\Sigma$ is a coupling constant, and $T_{p}$ denotes the phonon bath temperature. The coupling coefficient reads then $\Sigma=\frac{\pi^2 D^2\gamma_1 k_B^4}{60\rho \hbar^5v_F^3c^3} \sqrt{\frac{\gamma_1}{|\mu|}}$, where $D \simeq 30$ eV is the deformation potential, $\gamma_1 =0.4$ eV denotes the interlayer coupling, $\mu$ the chemical potential, $\rho=1.5 \cdot 10^{-6}$ kg m$^{-2}$ the mass density, $v_F = 1.0 \cdot 10^{6}$ m/s the Fermi velocity, and $c=2.6 \cdot 10^{4}$ m/s the speed of sound \cite{Efetov2010}. On the basis of gate sweeps on similar samples, we estimate the residual charge density  $\sim 5 \cdot 10^{11}$ cm$^{-2}$ which corresponds to chemical potential 17 meV. This yields $\Sigma = 0.12$ K$^4$Wm$^{-2}$, which has been verified experimentally in Ref.~\onlinecite{Yan2012}. At the largest electronic heating of $3.5$ nW  (Fig.~\ref{fig:cool}d),  the weak electron-phonon coupling results in $ T_{e}=10$ K.


The phonon temperature is bottlenecked by the Kapitza thermal boundary resistance between the graphene and gold. The present experiment allows for accurate measurement of this important quantity, not performed previously in dilution refrigerator temperatures. The data in Fig.~\ref{fig:cool}d is fitted as $\dot{Q}_{p-c} = A_c R_K  (T_{\m{p}}^4 - T_{\m{0}}^4)$, where the Kapitza conductivity $R_K = 1.3 \times 10^4$ WK$^{-4}$m$^{-2}$, and $A_c \simeq 9$ $\mu$m$^2$ is the contact area.

For 3d systems the following temperature dependence for the Kapitza conductivity is expected: $\dot{Q}_{p-c}= A_c R_{K} (T_{p}^4-T_{0}^4)$. In thin films, the density of phonon modes is different, and often an exponent between 3-4 is observed \cite{Swartz1989}. In graphene-metal systems, the temperature dependence of $R_K$ has been determined down to $\sim $ 50 K \cite{Schmidt2010,Balandin2011} but not below; similar values have been obtained for graphene-SiO$_2$ interfaces \cite{Chen2009} but there coupling to interfacial modes has been found to be important as well \cite{Freitag2009}. Consequently, only rather crude estimates are available: we employ the results of Refs.~\onlinecite{Schmidt2010,Chen2009}, extended by diffuse mismatch model \cite{ Duda2009} to the asymptotic region with $T^4$ dependence. This yields for Kapitza conductance values two orders of magnitude smaller than that obtained from fitting to Fig.~\ref{fig:cool}d but much larger than $\Sigma$. Hence, the graphene-substrate thermal contact seems better than expected from theory.



We now turn the discussion into the limitations and prospects of our scheme. In nearly all optomechanics experiments, high internal $Q$-value of the cavity is favored. Although the graphene-induced dissipation in the cavity has a negative effect for the internal $Q$, we note that the present demonstrations presumably do not suffer much from this. We believe that losses due to graphene can be avoided to some extent by careful impedance design of the cavity and graphene contacts, by introducing a possibility for tuning the graphene charge density, or by proximitizing the graphene to be superconducting.


It is intriguing to further estimate the prospect of ground-state cooling, quantified usually as $n_m < 1$, of the mechanical modes of graphene relying on microwave radiation pressure.
By increase of coupling $g$ one obtains, at a given bath temperature set by pump strength $n_c$, higher damping and thus more efficient cooling. We obtain that a coupling $g/2\pi \simeq 100$ Hz would be enough to cool to the ground state  in the present setting. In fact, the ideal coupling in the present case is not too far from here, however, the measured value was about three times smaller. We take the same reduction factor and thus we need an ideal $g/2\pi \simeq 300$ Hz. We suppose now a pristine monolayer graphene having the same dimensions as presently, but a narrower vacuum gap $d = 50$ nm. The cavity capacitance can further be reduced down to $C \sim 20$ fF by using low-dielectric substrate such as quartz \cite{MechAmpPaper}.
With these realistic values, we obtain the required coupling and excellent prospect to reach the quantum ground state of moving graphene. 

\emph{Acknowledgements} We thank Erno Damsk\"agg, Tero Heikkil\"a and Juha Pirkkalainen for useful discussions. This work was supported by the Academy of Finland, by the European Research Council (240387-NEMSQED) and by FP7 grant 323 924 iQUOEMS. The work benefited from the facilities at the Micronova Nanofabrication Center and at the Cryohall infrastructure. 

\bibliography{/Users/masillan/Documents/latex/MIKABIB}

\end{document}